\documentclass[twocolumn,prl,aps,superscriptaddress]{revtex4}

\usepackage{bm}
\usepackage{graphicx}
\usepackage{color}

\newcommand{\ybgg}{Yb$_{3}$Ga$_{5}$O$_{12}$}
\newcommand{\ggg}{Gd$_{3}$Ga$_{5}$O$_{12}$}

\begin{document}

\title{Uncommon magnetic ordering in the quantum magnet Yb$_{3}$Ga$_{5}$O$_{12}$} 

\author{S. Raymond\footnote{corresponding author : raymond@ill.fr}}
\affiliation{Univ. Grenoble Alpes, CEA, IRIG, MEM, MDN, 38000 Grenoble, France}
\author{E. Lhotel}
\affiliation{Institut N\'eel, CNRS \& Univ. Grenoble Alpes, 38000 Grenoble, France}
\author{E. Riordan}
\affiliation{Institut N\'eel, CNRS \& Univ. Grenoble Alpes, 38000 Grenoble, France}
\author{E. Ressouche}
\affiliation{Univ. Grenoble Alpes, CEA, IRIG, MEM, MDN, 38000 Grenoble, France}
\author{K. Beauvois}
\affiliation{Univ. Grenoble Alpes, CEA, IRIG, MEM, MDN, 38000 Grenoble, France}
\author{C. Marin}
\affiliation{Univ. Grenoble Alpes, Grenoble INP, CEA, IRIG, PHELIQS, 38000 Grenoble, France}
\author{M.E. Zhitomirsky}
\affiliation{Univ. Grenoble Alpes, Grenoble INP, CEA, IRIG, PHELIQS, 38000 Grenoble, France}

\date{\today}

\begin{abstract}
The antiferromagnetic structure of \ybgg~is identified by neutron diffraction experiments below the previously-known transition at $T_{\lambda}=54$ mK. The magnetic propagation vector is found to be ${\bf k}=(1/2, 1/2, 0)$, an unusual wave-vector in the garnet structure. The associated complex magnetic structure highlights the role of exchange interactions in a nearly isotropic system dominated by dipolar interactions and finds echos with exotic structures theoretically proposed. Reduced values of the ordered moments may indicate significant quantum fluctuations in this effective spin-1/2 geometrically frustrated magnet.
\end{abstract}

\maketitle
Research on geometrically frustrated magnetic materials has been extremely fruitful for discovering new states of condensed matter, namely exotic spin liquid, spin ice and spin fragmented states \cite{Balents, Jaubert, Lhotel1}. The hallmark geometries associated to these findings are the three dimensional pyrochlore and two dimensional kagome lattices. Recently interest has been reborn in the less studied hyperkagome lattice and in particular in the family of rare-earth gallium garnet systems $R_{3}$Ga$_{5}$O$_{12}$. In these cubic crystals (space group Ia$\bar{3}$d), the rare-earth ions $R^{3+}$, form two three dimensional interpenetrating lattices of corner sharing triangles called hyperkagome lattices. Depending on the rare earth element, anisotropy leads to contrasted behaviors from the realization of multi-axis antiferromagnetic order in Ising systems ($R=$Nd, Tb, Dy, Ho, Er) \cite{Hamman1,Hamman2,Hamman3,Cai,Wav,Petit,Kibalin,Hamman4} to exotic spin correlations in nearly isotropic systems $R=$Gd \cite{Paddison} and $R=$Yb \cite{Sandberg,Lhotel2}. 

In the present work, we focus on the magnetic state of the effective spin-1/2 hyperkagome material Yb$_{3}$Ga$_{5}$O$_{12}$. It has been known for a long time that the heat capacity exhibits a lambda-type anomaly at $T_{\lambda}  \approx 54$ mK and a broad hump at about 200 mK \cite{Filippi}. While the latter feature is associated with magnetic correlations recently studied by neutron scattering \cite{Sandberg,Lhotel2}, microscopic insight into the nature of the transition and the magnetic ground state has been missing for more than forty years. In particular, the M{\"o}ssbauer spectroscopy measurements did not observe any hyperfine splitting associated with a possible magnetic ordering below $T_{\lambda}$ \cite{Hodges}. The $\mu$SR experiments also did not detect any sign of long-range order as well pointing instead to the reinforcement of the dynamical correlations below $T_{\lambda}$ \cite{Dalmas03}. In this letter, we show through neutron diffraction experiments that a static long-range magnetic order does settle below $T_{\lambda}$. Our measurements reveal that it arises with a propagation vector ${\bf k}=(1/2, 1/2, 0)$ that has never been reported, to our knowledge, at zero field for the garnet systems, where the magnetic ordering occurs in almost all cases within the crystallographic unit cell, i.e. with a propagation vector $\textbf{k}=0$ ($\Gamma$ point of the Brillouin zone) \cite{garnet} and a few cases with $\textbf{k}= (0, 0, 1)$ ($H$ point of the Brillouin zone) \cite{Plumier}. 

Diffraction measurements were performed at the Institut Laue Langevin, Grenoble on the two-axis thermal-neutron diffractometer D23 equipped with a lifting detector. A copper monochromator provides an unpolarized beam with the wavelength of 1.276 {\AA}. The sample is the same as the one used for magnetization measurements in Ref. \onlinecite{Lhotel2}. In order to insure a good thermalization, the sample was inserted inside a sealed copper box filled with helium and the box was attached to the mixing chamber of a dilution stick. The based temperature was 30 mK on the thermometer mounted below the mixing chamber.

\begin{figure}
\includegraphics[width=9cm]{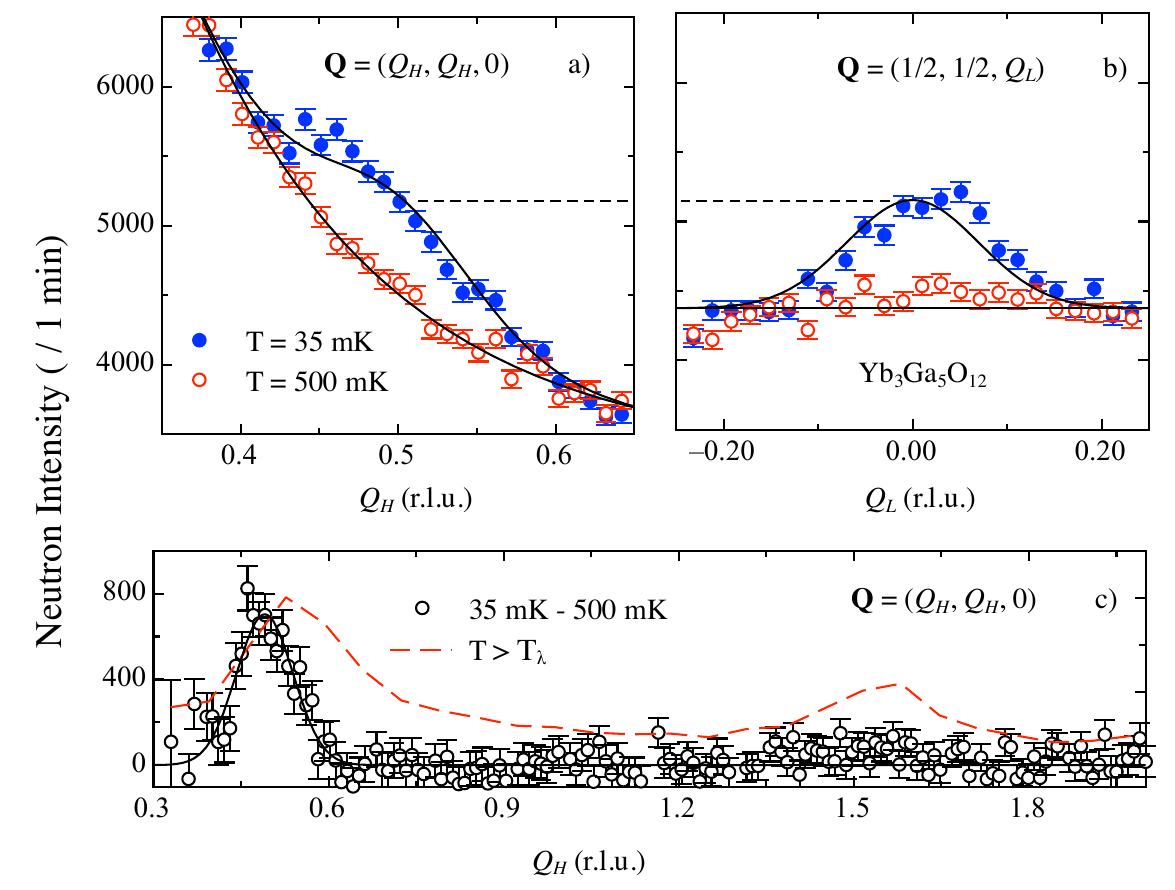}
\caption{\label{fig1} Momentum transfer cuts at 35 mK (full blue circles) and 500 mK (open red circles) along (a) ${\bf Q}=(Q_H, Q_H, 0)$ and (b) along ${\bf Q}=(1/2, 1/2, Q_L)$. (c) Subtraction of the neutron intensity obtained at 35 mK from the one obtained at 500 mK along ${\bf Q}=(Q_H, Q_H, 0)$ in an extended range up to $Q_H=2$. The red dashed line indicates the diffuse scattering data obtained above $T_{\lambda}$ in Ref. \onlinecite{Lhotel2}; the intensity is scaled to the new data. In all the panels, full lines are guides to the eye.}
\end{figure}

Based on common experimental observations for materials from the $R_{3}$Ga$_{5}$O$_{12}$ family as well as by theoretical predictions \cite{Capel,Kibalin}, an extensive search was performed for the magnetic ordering wave-vector ${\bf k}={\bf 0}$ complemented by a few checks at ${\bf k}=(0, 0, 1)$. No signal was found by subtraction of the peak intensity collected at 35 mK and 500 mK over 21 representative reflections (10 independent ones).
In contrast by expanding the search along a $\Gamma$-$\Gamma$ line, a magnetic signal has been detected.
Figures \ref{fig1}a and b show scans along the momentum transfer ${\bf Q}=(Q_H, Q_H, 0)$ and $(1/2, 1/2, Q_L)$ respectively at 35 mK and 500 mK, where the coordinates of the momentum transfer $\bf{Q}$ are given in units of $2\pi/a$ with the lattice parameter $a$ = 12.196 {\AA} (r.l.u. stands for reciprocal lattice units in the Figure \ref{fig1}). They reveal a peak at $Q_H=1/2$ and $Q_L=0$ respectively, so that the propagation vector is ${\bf k}=(1/2, 1/2, 0)$ (The relation between the momentum transfer ${\bf Q}$ and the propagation vector ${\bf k}$ is: ${\bf Q}={\bf G} \pm {\bf k}$ where $\bf{G}$ lies at a $\Gamma$ point of the reciprocal lattice).
Figure \ref{fig1}c shows the intensity difference obtained between the data collected at 35 and 500 mK along ${\bf Q}=(Q_H, Q_H, 0)$ over an extended range in $Q_{H}$. The diffuse scattering intensity measured above $T_{\lambda}$ on D7-ILL from Ref. \onlinecite{Lhotel2} is also shown as a dashed line. This highlights the fact that the correlations at $(1/2,1/2,0)$ evidenced above $T_{\lambda}$ in our previous study transform into a long range magnetic order below $T_{\lambda}$.

\begin{figure}
\includegraphics[width=7cm]{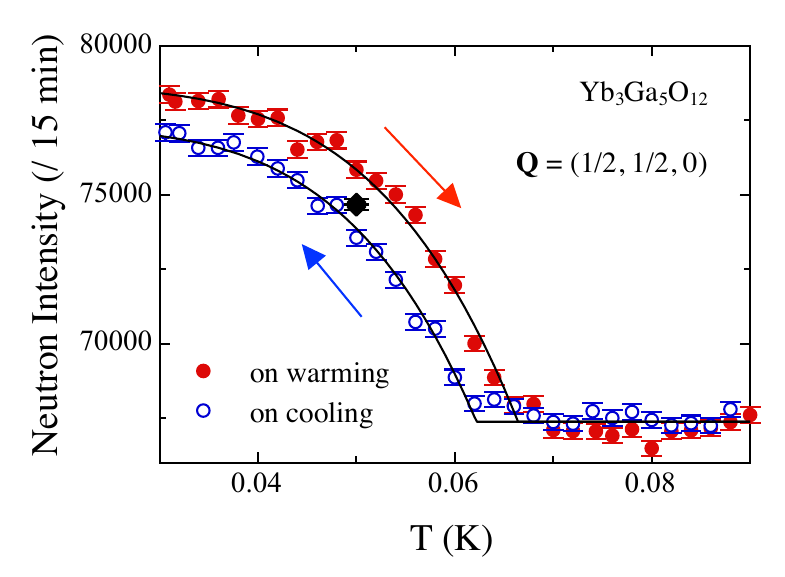}
\caption{\label{fig2} Temperature dependence of the neutron intensity measured at ${\bf Q}=(1/2, 1/2, 0)$ on warming from 30 to 90 mK (red dots) and on cooling from 90 to 30 mK (blue circles). The black diamond points correspond to data obtained by longer stabilisation time of 1 h (see text).}
\end{figure}

Figure \ref{fig2} shows the temperature dependence of the Bragg peak intensity at ${\bf Q}=(1/2, 1/2, 0)$ on warming and on cooling. The temperature was varied by 2 mK steps, and after a 4 min waiting time, the intensity was counted during 15 min for each point. 
  The obtained transition temperatures are 62 mK on cooling and 66~mK on warming, which may suggest a first order transition. However, after cooling the initial low temperature intensity is not recovered. In order to check whether it is intrinsic or related to the thermalization of the sample, a point at 50 mK has been measured with a longer stabilization time. The results are presented in Figure \ref{fig2} where the black diamond points indicate the data obtained at 50~mK after 1 hour waiting time on warming from 30 mK or 1 hour waiting time on cooling from 90~mK. The same intensity is recovered (overlapping diamonds) meaning that there is no intrinsic temperature hysteresis in the system. Due to the available neutron beamtime, we could not determine the equilibrium temperature dependence of the magnetic intensity with this stabilization time protocole for each point. From this analysis, we estimate that the real N\'eel temperature is $T_N$= 64 $\pm$ 2  mK, a somewhat higher value than the reported $T_{\lambda}$=54 mK in specific heat. 

\begin{figure}
\includegraphics[width=9cm]{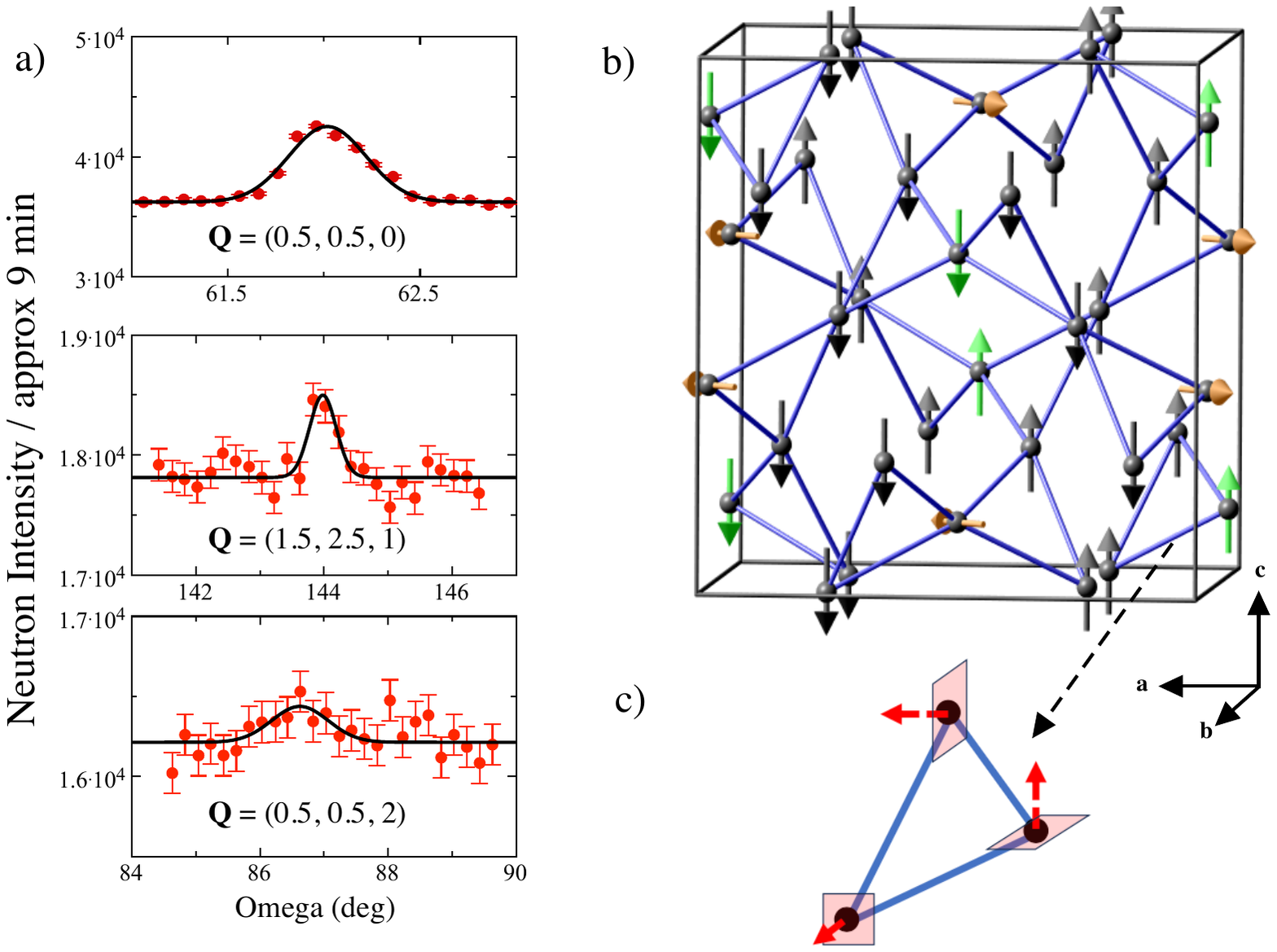}
\caption{\label{fig3} a) Representative rocking curves (red points) performed at 35 mK for ${\bf Q}=(0.5, 0.5, 0)$, ${\bf Q}=(1.5, 2.5, 1)$ and ${\bf Q}=(0.5, 0.5, 2)$ and the associated fitted intensity using a Gaussian function (black lines). b) Magnetic structure in the cubic crystallographic cell (1a sites in black, 1b sites in green, 1c sites in orange).  c) Zoom on a triangle showing the local axes in red with $z$ shown as an arrow and the ($x$, $y$) plane depicted as a parallelepiped.}
\end{figure}

In the space group Ia$\bar{3}$d, the star of ${\bf k}=(1/2, 1/2, 0)$ contains six equivalent wave-vectors, producing six antiferromagnetic domains. As expected in the absence of symmetry breaking perturbation, the integrated peak intensities measured for the six corresponding Bragg reflections ${\bf Q}={\bf k}$ are found to be very similar.
In order to address the magnetic structure, a collection of magnetic peaks belonging to the single configuration domain with ${\bf k}=(1/2, 1/2, 0)$ has been carried out through rocking curve measurements (``Omega scans"). Representatives of such scans are shown in Figure \ref{fig3}a, the strongest peak, which corresponds to ${\bf Q}={\bf k}$ being shown in the top panel. Due to the weak signal, long counting times were necessary (9 min per point) and 21 independent magnetic Bragg peaks could be collected.
Owing to the limited set of data and given the complexity of the lattice (12 magnetic sites without the centring I), the full structure could not be solved. 
The discrimination between the possible structures was made by least-square fitting of the data using the basis vectors given by the magnetic representation analysis (Details are given in the Supplemental Material \cite{Supplement}).

In the space group Ia$\bar{3}$d with the propagation vector ${\bf k}=(1/2, 1/2, 0)$, there exist only two representations, $\Gamma_1$ and $\Gamma_2$, and the sites are split into 1a sites (8 moments), 1b sites (2 moments) and 1c sites (2 moments) \cite{Supplement}. The remaining 12 sites are deduced by the centring symmetry I, where the moments are reversed given the propagation vector $\bf{k}$.
The refinement indicates that the structure has the $\Gamma_2$ representation symmetry and that the 1a site moments are along (0, 0, $M_Z$) (the components ($M_X$, $M_Y$, $M_Z$) are given in the global coordinate frame ($X$, $Y$, $Z$) associated to the crystallographic $a$, $b$ and $c$ axes).
As far as the 1b site is concerned, the fits show that a $M_Z$ component must exist but they are not conclusive about the symmetry allowed ($M_X$, -$M_X$, 0) component.
For the 1c site, the data are compatible with any possibility from zero magnetic moment to magnetic moments in the direction ($M_X$, -$M_X$, $M_Z$).

\begin{table}[b!] 
\caption{\label{table} Spin components for the proposed structure of Yb$_{3}$Ga$_{5}$O$_{12}$. The magnetic moment components are given in the global coordinate frame (($M_X$, $M_Y$, $M_Z$) notation) and in the local coordinate frame (($m_x$, $m_y$, $m_z$) notation) at each site indicated by its crystallographic coordinates in units of $a$. The relations between the local and global coordinate frames are given in the Supplemental Material \cite{Supplement}.}

\begin{ruledtabular}
\begin{tabular}{cccccc}
Site &  coordinates &  $M$ global frame & $m$ local frame \\
\hline
1a & $(1/8, 0, 1/4)$	& $(0, 0, M_Z)$	& $(-m_x, m_x, 0)$ \\
     & $(3/8, 0, 3/4)$	& $(0, 0, M_Z)$ 	& $(-m_x, m_x, 0)$ \\
     & $(1/4, 1/8, 0)$	& $(0, 0, M_Z)$ 	& $(m_x, m_x, 0)$ \\ 
     & $(3/4, 3/8, 0)$	& $(0, 0, -M_Z)$ & $(m_x, m_x, 0)$  \\ 
     & $(7/8, 0, 3/4)$	& $(0, 0, M_Z)$ 	& $(-m_x, m_x, 0)$  \\
     & $(5/8, 0, 1/4)$	& $(0, 0, M_Z)$ 	& $(-m_x, m_x, 0)$  \\
     & $(3/4, 7/8, 0)$	& $(0, 0, -M_Z)$ 	& $(-m_x, -m_x, 0)$ \\ 
     & $(1/4, 5/8, 0)$	& $(0, 0, M_Z)$  	& $(-m_x, -m_x, 0)$  \\   
\hline
1b & $(0, 1/4, 1/8)$	& $(0,0,M_Z)$ 	& $(0,0,m_z)$\\
     & $(0, 3/4, 7/8)$	& $(0,0,M_Z)$ 	& $(0,0,m_z)$  \\
\hline
1c  & $(0, 3/4, 3/8)$	& $ (-M_X,M_X,0)$ 	& $(0,m_y,0)$  \\
      & $(0, 1/4, 5/8)$	& $(-M_X,M_X,0)$ 	& $(0,m_y,0)$ \\ 
\end{tabular}
\end{ruledtabular}
\end{table}

In order to go further it is thus necessary to consider the different terms of the Hamiltonian of the system, i.e. the anisotropy and the interactions that are at play. We have recently shown that a dipolar model reproduces very well two experimental results in Yb$_{3}$Ga$_{5}$O$_{12}$, the positive Curie-Weiss temperature $\theta_p=97$ mK and the weakly dispersive magnetic excitation spectrum in the field saturated paramagnetic phase \cite{Lhotel2}. The obtained magnetic structure should thus minimize the energy associated to dipolar interactions. Concerning the anisotropy, the crystal electric field (CEF) leads to the ground-state doublet, well separated from the first excited state (the splitting is about 67 meV) \cite{Carson, Buchanan, Pearson}. As a result, at low temperatures the excited crystal-field levels can be ignored and Yb$^{3+}$ ions
are faithfully represented by effective spins $S=1/2$. In the cubic unit cell, 24 ytterbium sites are split into six groups according to the local crystal-field axes ($x$, $y$, $z$), depending on the orientation of the oxygen dodecahedron around each site (the relations between the local and global coordinate frames are given in the Supplemental Material \cite{Supplement} and the local axes are shown in Figure \ref{fig3}c). The $g$-tensor for the ground-state doublet is described by values $g_x=3.74$, $g_y=3.60$ and $g_z=2.85$ in the local coordinate frame \cite{conventions}. Hence, the doublet is fairly isotropic with a weak easy-plane anisotropy. This anisotropy is expected to frustrate the system with respect to collinear ordering, and is thus foreseen to play an important role in the observed magnetic structure.

The combined effect of dipolar interactions, exchange interactions and anisotropy has been theoretically addressed in the early studies of gallium garnets \cite{Capel} and reexamined in a recent investigation \cite{Kibalin}. Such studies are nevertheless limited to ${\bf k}={\bf 0}$ corresponding to an intra-cell magnetic ordering of the 24 magnetic atoms. 
For Ising systems $g_z \gg g_x,\ g_y$, antiferromagnetic ordering occurs in a multi-axial structure with moments along the local axis $z$ (so-called ``AFA states"). For a given triangle, the local $z$-axis alternates between the three $X$, $Y$ and $Z$ global axes. This solution is given by a single basis vector in magnetic representation analysis (There are two such representations, odd or even with respect to the inversion). These structures occur for $R=$Nd, Tb, Dy, Ho, Er \cite{Hamman1,Hamman2,Hamman3,Cai,Wav,Petit,Kibalin,Hamman4}.  In the case of nearly isotropic systems such as Yb garnet, a more complex ferrimagnetic structure was proposed, the so-called ``FC state", arising from a three dimensional representation with nine basis vectors \cite{Capel}. Note that this ferrimagnetic FC state is associated with an odd representation with respect to the inversion and an even representation also exists leading to a complex antiferromagnetic state  (the odd/even classification is only valid for $\bf{k=0}$). 

Interestingly, \ybgg\ exhibits magnetic order at ${\bf k}=(1/2, 1/2, 0)$, which has not been anticipated so far in theoretical studies. Based on the results obtained for ${\bf k}={\bf 0}$, we propose the magnetic structure detailed in Table \ref{table}: as explained above, to be consistent with the measured intensities, the magnetic moments 1a have to be along (0, 0, $M_Z$), which actually lies in the easy local anisotropy planes of these ions, i.e with component $(m_x, m_y, 0)$. Such a collinear state has similarities with one possible FC state introduced in Ref.~\onlinecite{Kibalin}. For 1c magnetic moments, the refinement only imposes the $(-M_X, M_X, M_Z)$ constraint, and it thus appears logical to put these moments in their easy-plane. This amounts to cancel the $Z$ component, giving a $(0, m_y, 0)$ moment in the local frame. Note that this orientation cancels the nearest neighbor Heisenberg exchange interaction since the nearest neighboring 1a sites have moments along $Z$. However an absence of ordered magnetic moment on the 1c site cannot be ruled out with our data. Theoretical studies on Ising spins on the hyperkagome lattice in the presence of long-range interactions could stabilize such magnetic configurations with paramagnetic sites ($\bf{k}$=0 calculation) \cite{Yoshioka}, but we can expect that the less constrained anisotropy in \ybgg\ will favor the presence of an ordered moment as supposed here. 
Finally, for the 1b moments, a non-zero $M_Z$ component (which corresponds to a $m_z$ component in the local frame), which opposes to the local anisotropy, is obtained from the refinement. We propose that the other components (whose values do not impact the quality of the refinement) are zero, so that, in analogy with the FC state introduced above, the 1b moments are along $(0, 0, M_Z)$ and collinear to the 1a ones.

\begin{figure}
\includegraphics[width=8.5cm]{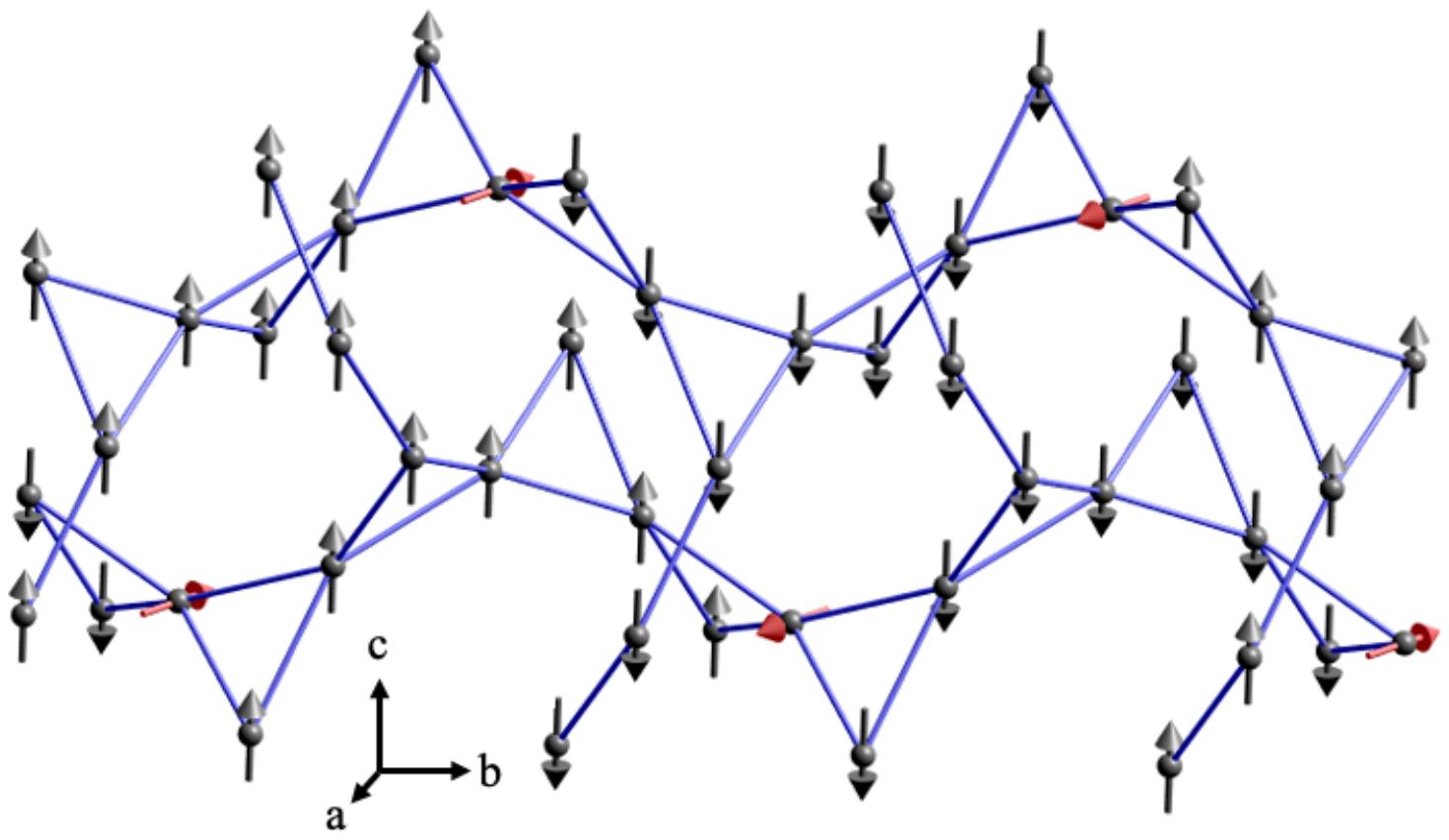}
\caption{\label{fig4} Fragments of the magnetic structure highlighting the underlying organization: the spins of the 1c sites (red arrows) separate lines / blocks of up and down spins (black arrows). The shown cut corresponds to $-0.25a~\leq~X~\leq~0.4a$, $0~\leq~Y \leq~2.3b$ and $-0.5c~\leq~Z \leq~0.5c$.}
\end{figure}

The final magnetic structure is shown in Figure \ref{fig3}b. To better visualize it, fragments of the structure are displayed in Figure \ref{fig4}. This view indeed points out the fact that the 1c sites separate lines / blocks of spin up and spin down magnetic moments from 1a and 1b sites: the 1c site is shared by two triangles, one with its two remaining spins along $(0, 0, M_Z)$ and the other with its spins along the opposite directions $(0, 0, -M_Z)$.
Such arrays of parallel spins could be favoured by dipolar interactions, that are expected to be dominant in \ybgg\ \cite{Lhotel2}.

The obtained ordered moment associated to this magnetic structure is 0.61 (1)~$\mu_{\rm B}$. We have constrained its magnitude to be equal on each site and taken into account the 6 configuration domains while assuming equal population of the 8 orientation domains. The ordered moments at $T= 30$ mK account for only about 1/3 of the saturated moments obtained in the magnetization measurements \cite{Lhotel2}. 
Normally, one would expect to find $\approx$ 3/4 of the full moments at a temperature roughly twice smaller than $T_N$. It is to be noted that this value of the ordered magnetic moment is robust when assuming zero magnetic moment on the 1c site. Two physical mechanisms behind the moment reduction in \ybgg~can be considered. First, geometric frustration on the hyperkagome lattice promotes significant thermal fluctuations at $T<T_N$ as was discussed previously for the Ising antiferromagnet Ho$_{3}$Ga$_{5}$O$_{12}$ \cite{Zhou}. Second, an even more interesting possibility is that strong quantum fluctuations for nearly isotropic spin-1/2 Yb moments reduce magnetic order even at $T = 0$ K. For comparison with other three-dimensional frustrated spin models, note that 30\%\ moment reduction was found for the nearest-neighbor Heisenberg spin-1/2 antiferromagnet on a fcc lattice \cite{Schick22}, whereas even larger quantum effects are expected for anisotropic spin-1/2 pyrochlore antiferromagnets in the proximity to classical boundaries \cite{Zhitomirsky12,Yan17}. Inelastic neutron scattering at first sight appears as a promising technique to investigate the fluctuations in the ordered phase, but it is worth noting that the experimental difficulties in thermalizing \ybgg\ samples below 60 mK, while keeping reasonable background, would make these measurements on large single crystals below the transition very challenging.

We now address the puzzling result of the propagation vector ${\bf k}=(1/2, 1/2, 0)$, which has never been reported to our knowledge for the magnetic structure of rare-earth garnets in zero field, where ordering with ${\bf k=0}$ is the paradigm. From our analysis, dipolar interactions and weak anisotropy appear to be the most important ingredients, but they are expected to give ${\bf k}={\bf 0}$ ordering. 
Additional antiferromagnetic interactions are thus needed to stabilize the unusual propagation vector evidenced in the present neutron diffraction experiments. It is worth noting that the role of exchange interactions in \ybgg\ remains debated from previous studies above the transition temperature \cite{Sandberg,Lhotel2}. In Ref. \onlinecite{Sandberg}, two Ising models with dominant nearest neighbor interactions are considered to reproduce distinct features of the diffuse scattering arising at different momentum transfer moduli. Interestingly, the introduction of a next nearest neighbor interaction allows the authors to describe the features at ``high"  $Q$, but fails in reproducing the low $Q$ features. The energy scale obtained for the first neighbor interaction is nevertheless not compatible with the field induced properties reported in Ref. \onlinecite{Lhotel2}. In the latter, a purely dipolar model is introduced and reproduces well the Curie-Weiss temperature and the magnetic excitation spectrum energy range in the field saturated phase. Nevertheless, the width of the excitation spectrum suggested the existence of subleading exchange interaction terms. The magnetic structure found in the present work seems to confirm the latter scenario. The parallel arrangement of most neighbouring spins in a direction essentially determined by the local anisotropy is indeed in favor of dominant dipolar interactions, all the more the obtained structure ressembles a ${\bf k=0}$ predicted pattern using a dipolar model \cite{Kibalin}. In addition, subleading antiferromagnetic exchange interactions, probably up to several neighbors, must be present to explain the peculiar ordering propogation vector ${\bf k}=(1/2, 1/2, 0)$. New inelastic neutron scattering studies were recently performed in order to gain insight into the hierachy of interactions in \ybgg\ \cite{Deen24}.

It is interesting to compare \ybgg~with the extensively studied sister material \ggg. Gadolinium ions have large semiclassical spins $S$=7/2 with a small easy-plane anisotropy  in \ggg ~\cite{Paddison, Bonville}. No magnetic transition is known for this compound, though complex spin-spin correlations based on ten-site loops were reported that give rise to a so-called director state \cite{Paddison}. A partial incommensurate ordering was also evidenced \cite{Petrenko98}, as well as spin-glass like freezing \cite{Schiffer}. These properties are the consequences of competing small energy scales. It was shown that two types of next-nearest neighbor interactions  have to be included in the Hamiltonian to understand the observed properties even if they are much smaller than the dominant antiferromagnetic nearest neighbor interactions and dipolar interactions \cite{Yavorskii}. 
In \ggg, for a magnetic field ${\bf H}$ applied parallel to $[1, \bar{1}, 0]$, a ${\bf k}=(1/2,1/2,0)$ ordering vector has been reported in a narrow field range, included in a larger region of the phase diagram with four incommensurate wave-vectors \cite{Petrenko}, which highlights the fragility of such a phase. Calculations show that, specifically for this field direction, soft modes leading to long range order are possible along the line $(h, 1-h, 0)$, which includes (1/2, 1/2, 0) \cite{Ancliff}. This comparison with \ggg\ suggests that in \ybgg, the ${\bf k}=(1/2, 1/2, 0)$ ordered phase results from a delicate balance between several next nearest exchange interactions that affect the expected ground state expected for dipolar interactions only.

In the present work, we have considered a single-${\bf k}$ multi-domain structure but a multi-${\bf k}$ structure cannot be ruled out. Multi-${\bf k}$ structures are often difficult to distinguish from single-${\bf k}$ ones by neutron diffraction experiments. Remarkably, in the pyrochlore lattice system Gd$_{2}$Ti$_{2}$O$_{7}$, a  2-$\bf{k}$ magnetic structure is now established \cite{Paddison2} with a ${\bf k}=(1/2, 1/2, 1/2)$ propagation vector, while similarly to garnets, most magnetically ordered pyrochlores  have a {\bf k}={\bf 0} propagation vector. Other multi-${\bf k}$ structures occurring for magnetic ordering at half integer reciprocal lattice vector were also theoretically shown to be stabilized in cubic spinels \cite{Zhito}. Since the proposed single-${\bf k}$ structure is almost collinear, further inelastic neutron scattering experiments aiming to determine the polarization of the lowest energy excitations would indicate if a more complex multi-${\bf k}$ structure should be considered \cite{Jensen}.

To conclude, more than forty years after the discovery of the lambda anomaly at 54 mK in Yb$_{3}$Ga$_{5}$O$_{12}$, long range static antiferromagnetic ordering is revealed with the unusual propagation vector ${\bf k}=(1/2, 1/2, 0)$ for the garnet lattice. The associated magnetic structure has similarities with complex spin arrangements already theoretically proposed for $\bf{k}$=0 and never experimentally observed so far, even for this latter propagation vector.  This emphasizes the delicate balance between the competing energy scales at play in these systems in order to realize the unique magnetic ordering pattern revealed in \ybgg. Additionally, the reduced ordered moment points to significant fluctuations, whose origin is to be addressed by further experimental and theoretical investigations. Together with the different ground state achieved in the notorious system \ggg, these results highlight the rich possibilities brought by the hyperkagome lattice in the limit of small anisotropy and the importance of this geometry in the overall progress in undestanding frustrated magnetism.

\bigskip
We acknowledge B. Vettard, P. Fouilloux, M.-H Baurand, V. Joyet, E. Bourgeat Lami and X. Tonon for technical support. 
We acknowledge S. Petit and F. Damay for fruitful discussions.
The structure was determined using Mag2POL \cite{Mag2POL} and checked with Fullprof \cite{Fullprof}.
We acknowledge help from N. Qureshi for Mag2POL.
E. L., C. M., and M. E. Z. acknowledge financial support from Agence Nationale de la Recherche, France, Grant No. ANR-18-CE05-0023.
The neutron scattering data collected at the ILL for the present work are available at Ref. \cite{DOI1}.


\begin{thebibliography}{99}
\bibitem{conventions} The conventions are like in e.g. Ref. \cite{Kibalin} with $z$ a quaternary cubic axis and $x$ and $y$ binary axes. It is different to the one used in e.g. Ref. \cite{Paddison} where $x$ is the quaternary axis and $y$, $z$ are binary axes.
\end{thebibliography}
\end{document}